# Complete and robust energy conversion by sum frequency generation based on Invariant Engineering

Cong-Fu Zhang[1,3], Zhao-Lu Wang[1], and Hong-Jun Liu[1,2]

[1]State Key Laboratory of Transient Optics and Photonics, Xi'an Institute of Optics and Precision Mechanics, Chinese Academy of Science, Xi'an 710119, China
[2]Collabotative Innovation Center of Extreme Optics, Shanxi University, Taiyuan 030006, China
[3]University of Chinese Academy of Sciences, Beijing 100084, China

**We propose an analytical method to achieve complete energy conversion in sum frequency generation based on Lewis-Riesenfeld invariants theory. This technique, derived from a two-level atom transition in quantum mechanics, is more efficient and robust than conventional methods. In our scheme, the quasi-adiabatic single control parameter model is established, and the value of single control parameter is selected to make the initial eigenstate perfectly converted to the final eigenstate we need. Corresponds to the nonlinear frequency conversion process, the nonlinear crystal structure is designed with the inverse engineering of optimal control theory, which is robust against the perturbations in the coupling coefficient and phase mismatch, including pump intensity and crystal polarization period variations, and results in almost 100% conversion efficiency at any crystal length. It is demonstrated that the frequency conversion can be achieved in the wavelength range of 2.6 μm -3.6 μm with a spectral bandwidth of the conversion efficiency over 50% approaching to 400 nm when the crystal length $L=1$ mm.**

## 1 Introduction

Optical frequency conversion is a phenomenon that the nonlinear response of light waves with different frequencies in nonlinear materials produces light with different third frequencies, which has broad application prospects in quantum optics, new light sources, holographic imaging, biological imaging and sensing and other fields[1]. Generally speaking, the highly efficient frequency conversion process

Hong-Jun Liu: liuhongjun@opt.ac.cn

needs to satisfy the phase matching condition or quasi-phase matching condition, so that the phase matching $\Delta k=0$. Phase matching is the manifestation of momentum conservation in the process of nonlinear light-wave interaction, which directly determines the efficiency of nonlinear frequency conversion. When the phase mismatch is large, the conversion efficiency is very low. There are several ways to adjust the birefringence by adjusting the angle of propagation of the wave through the nonlinear crystal or the temperature of the crystal, or to compensate for the phase mismatch of the wave vector by changing the polarization period of the crystal. Unfortunately, neither of above standard frequency conversion technologies can provide simultaneously highly efficient and broadband frequency conversion, which is of great practical importance for many applications.

In the last decade, Suchowski et al. have theoretically proposed a completely different efficient, robust and broadband frequency conversion method, named adiabatic frequency conversion technique, by comparing the frequency conversion process to the dynamics of a two-level atomic system Landau-Zener(LZ) theorem based on undepleted pump approximation and have demonstrated this method experimentally by the chirped polarized nonlinear crystals are designed without phase matching [2-8]. In addition, Andery Markov et al.[9] show that it is feasible to achieve adiabatic process by setting the temperature gradient in the crystal; Xun Liu et al.[10]proposed an adiabatic conversion method based on electro-optic effects. All of these methods illustrate the nature of adiabatic theory: it is a process requiring long distance evolution to achieve efficient frequency conversion, i.e., the adiabatic conditions strictly limit the phase mismatch changing very slowly from very large negative values to very large positive values and requiring high pumping intensity [11].

In the other hand, "shortcuts to adiabaticity" (STA) method [12-15], providing a physical model in quantum control theory to speed up the above quantum adiabatic slow process by a fast path, the final result of slow adiabatic evolution can also be obtained at the same time. The adiabatic scheme designed based on quantum adiabatic shortcut technology includes three methods: transition-less quantum driving [16], super-adiabatic method (LR) quantum invariants [17] and inverse engineering based on Lewis-Riesenfeld [18]. Especially, Lewis-Riesenfeld invariant theory, combining time-dependent perturbation theory and optimal control theory [19, 20], reduces the influence of different perturbations on the system evolution and increases system robustness against error noise. This technique is beyond the conventional adiabatic limits, and is now commonly used to design different types of efficient and robust photonic waveguide devices based on the analogies between quantum mechanics and wave optics, such as adiabatic couplers [21-27] and polarization conversion devices[28], which has a shorter evolutionary length than Landau-Zener adiabatic designs and can achieve 100% conversion efficiency.

In this paper, we propose a quasi-adiabatic (QA) single control parameter model with the application of Lewis-Riesenfeld invariants, which can achieve STA during frequency upconversion by designing the coupling coefficient and structure of nonlinear crystal polarization period. In our model, the coupling coefficient is constant and the phase mismatch is a variable related to the coupling coefficient and propagation distance $z$. By combining the Lewis-Riesenfeld invariants theory with the optimal control theory[19, 20], the optimal schemes against the perturbation of coupling coefficient perturbation and phase mismatch can be designed respectively. These optimal schemes allow us to calculate the optimal solution of the coupling coefficient corresponding to any crystal length, thus the design of the crystal polarization period by this scheme can achieve a conversion efficiency of almost 100%. Further, considering the crystal length $L$ =1 mm as an example for numerical calculation, it shows that our scheme can provide efficient broadband sum frequency generation in the 2.6 μm - 3.6 μm wavelength range with a spectral bandwidth approaching to 400 nm bandwidth. Compared with the conventional invariant method, the quasi-adiabatic model has a smaller coupling coefficient, which means that complete energy conversion can be achieved with a lower pump intensity. The numerical results show that our optimal schemes are robust against the influence of the pump intensity variations and the crystal polarization periodic processing error. Finally, the comparation between Landau-Zener adiabatic frequency conversion and quasi-adiabatic method is carried out, the advantages of our method are demonstrated. We prove that the scheme based on Lewis-Riesenfeld invariant engineering can keep complete and robust energy conversion by sum frequency generation at any crystal length in weak signal upconversion.

## 2 Quasi-adiabatic single control parameter model

We consider an upconversion process in which two light fields of different angular frequencies ($\omega_1$ and $\omega_2$) interact inside a nonlinear crystal to produce a light field of $\omega_3$, where 1, 2, and 3 are signal light, pump light and upconversion light respectively, satisfying the conditions $\omega_1 + \omega_2 = \omega_3$. If the pump intensity is much larger than the signal light, it can be regarded as no loss in the propagation process, which means that the undepleted pump approximation is applied, the pump amplitude $A_2$, can be considered to be constant along the propagation. Then, considering the plane wave approximation condition and ignoring the group velocity dispersion, the three-wave coupling process of frequency upconversion can be written in the form of linear equations[2, 3],

$$\begin{cases} i\dfrac{\partial \tilde{A}_1(z)}{dz} = \kappa \tilde{A}_3 e^{-i\Delta k z} \\ i\dfrac{\partial \tilde{A}_3(z)}{dz} = \kappa \tilde{A}_1 e^{i\Delta k z} \end{cases}. \quad (1)$$

Here, $z$ is the position along the propagation axis, $\Delta k = k_1 + k_2 - k_3 + k_\Lambda$ and $k_i = \dfrac{2 n_i \pi}{\lambda_i}$ with $i$ =1, 2, 3 is signal, pump, and upconversion wave respectively, $k_\Lambda = \frac{2\pi}{\Lambda}$, $\Lambda$ is the polarization period, and the $\kappa = \frac{4\pi\omega_1\omega_3}{c^2\sqrt{k_1 k_3}}\chi_1 A_2$ is the coupling coefficient, $\chi_1 = \frac{2}{\pi}\sin(\pi D)\chi^{(2)}$, $D$ is the duty cycle and $\chi^{(2)}$ is the second order nonlinear susceptibility of the nonlinear crystal, and the normalized signal, pump, and upconversion amplitudes are $\tilde{A}_1$, $A_2$ and $\tilde{A}_3$.

Therefore, the sum frequency process can be written in matrix form $\frac{\partial \mathbf{A}}{dz} = -i\mathbf{H}_0(z)\mathbf{A}$, $\mathbf{A} = [A_1 \ A_3]^T$, $A_1 = \tilde{A}_1 e^{i\Delta k z/2}, A_3 = \tilde{A}_3 e^{-i\Delta k z/2}$, and

$$\mathbf{H}_0(z) = \frac{1}{2}\begin{bmatrix} -\Delta k & \kappa \\ \kappa & \Delta k \end{bmatrix}. \quad (2)$$

It is shown that Eq (2) is formally equivalent to the time-dependent Schrödinger equation, so the frequency conversion between coupled waves is similar to the population transfer between two-level atom. Next, we will design the coupling constant and phase mismatch $\Delta k$ through the combination of invariant-based inverse engineering and perturbation theory to achieve an efficient and robust sum frequency process.

According to the Hamiltonian $H_0$ of the sum frequency system given above, the dynamic invariant $I(z)$ is[18]

$$\mathbf{I}(z) = \frac{1}{2}\Omega_0 \begin{pmatrix} \cos\theta & \sin\theta e^{-i\beta} \\ \sin\theta e^{i\beta} & -\cos\theta \end{pmatrix}, \quad (3)$$

where $\theta \equiv \theta(z)$ and $\beta \equiv \beta(z)$ are $z$-dependent angles, $\Omega_0$ is an arbitrary constant. The eigenstates of the invariant $I(z)$ have the same form as that of $H_0$, which can be expressed as

$$\begin{cases} |\phi_1(z)\rangle = \begin{pmatrix} \cos\frac{\theta}{2}e^{-i\beta} \\ \sin\frac{\theta}{2} \end{pmatrix} \\ |\phi_2(z)\rangle = \begin{pmatrix} \sin\frac{\theta}{2} \\ -\cos\frac{\theta}{2}e^{i\beta} \end{pmatrix} \end{cases}, \quad (4)$$

with eigenvalues $\lambda_\pm = \pm\Omega_0/2$.

By constructing the Lewis-Riesenfeld invariant theory, dynamic invariants can be used to represent arbitrary solution of the time-dependent Schrödinger equation, in general the system state can be written as

$$|\Psi(z)\rangle = \sum_n c_n e^{i\alpha_n(z)}|\phi_n(z)\rangle, \quad (5)$$

where the $c_n$ are $z$-independent amplitudes and $|\phi_n(z)\rangle$ is the eigenvector of $I(z)$. Lewis-Riesenfeld invariant phase is expressed as

$$\alpha_\pm(z) = \frac{1}{2}\int \left(\dot\beta + \frac{\dot\theta\cot\beta}{\sin\theta}\right)dz', \quad (6)$$

$\theta \equiv \theta(z)$ and $\beta \equiv \beta(z)$ can be given by the invariant definition $\frac{\partial \mathbf{I}}{\partial t} + \frac{1}{i\hbar}[\mathbf{I}, \mathbf{H}_0] = 0$ [18],

$$\dot\theta(z) = -\kappa(z)\sin\beta(z), \quad (7)$$
$$\dot\beta(z) = -\kappa(z)\cot\theta(z)\cos\beta(z) - \Delta k(z). \quad (8)$$

The following is a discussion on the value of coupling coefficient $\kappa$. On the one hand, since the conventional Lewis-Riesenfeld invariant is proposed in the two-level atomic transition, and the Hamiltonian of the two-level system is[29]

$$\mathbf{H}_a(t) = \frac{\hbar}{2}\begin{bmatrix} -\Delta(t) & \Omega(t) \\ \Omega(t) & \Delta(t) \end{bmatrix}, \quad (9)$$

where $\Omega(t)$ represents the Rabi frequency of the outer field laser corresponding to the coupling coefficient $\kappa$ in Eq. (2). Generally speaking, the Rabi frequency is regarded as a time dependent pulse function. In fact, if the energy total of the laser ($W = \int_0^T \Omega(t)dt$) is the same, then the energy level transition can be achieved as well, that is, the Rabi frequency and coupling coefficient can be set to a constant.

On the other hand, if $\kappa$ varies with the propagation distance $z$, observing the expression $\kappa = \frac{4\pi\omega_1\omega_3}{c^2\sqrt{k_1 k_3}}\chi_1 A_2$ with $\chi_1 = \frac{2}{\pi}\sin(\pi D)\chi^{(2)}$, in this way the duty cycle in nonlinear susceptibility is a function of the spatial variation of the propagation distance $z$. While for nonlinear crystals, it is very difficult to realize the elaborate processing of duty cycle and polarization period in a short crystal simultaneously. More importantly, the pump intensity of the conventional invariant scheme is so high that exceed the damage threshold of the crystal, which will be explained in detail below.

Therefore, we set the coupling coefficient is a constant, and the phase mismatch $\Delta k$ is regarded as a variable varying with the propagation distance $z$ in our scheme, that is, the polarization period of the crystal is variable with the duty cycle $D = 0.5$. By this design, we obtain a design scheme whose crystal structure is similar to that of adiabatic chirped crystal with a lower pump intensity, which is called quasi-adiabatic model.

Then we take the derivative of Eq. (7) with respect to $z$:

$$\ddot{\theta}(z) = -\kappa \dot{\beta}(z) \cos \beta(z), \quad (10)$$

substituting Eqs. (7) and (10) into Eq. (8), a new form for the phase mismatch can be obtained:

$$\Delta k(z) = \frac{\ddot{\theta}(z)}{\kappa \sqrt{1 - \left(\frac{\dot{\theta}(z)}{-\kappa}\right)^2}} - \kappa \cot \theta(z) \sqrt{1 - \left(\frac{\dot{\theta}(z)}{-\kappa}\right)^2} \quad (11)$$

This means that for different coupling coefficients, corresponding $\Delta k$ evolution schemes can be found to satisfy our requirements. The trajectory of state evolution between them can be parametrized according to one of the eigenstates $|\phi_1(z)\rangle$ of the invariant. In order to achieve complete energy conversion between signal light and upconversion light, by using $|\phi_1(z)\rangle$ in Eq. (4), the boundary conditions of the initial and final states can be set as

$$|1\rangle = \begin{pmatrix} 1 \\ 0 \end{pmatrix}, |2\rangle = \begin{pmatrix} 0 \\ 1 \end{pmatrix}. \quad (12)$$

In the invariant-based engineering, since the system evolves strictly according to Eq. (4), we can choose arbitrary interpolation function $\theta$ and use inverse engineering to design system parameters as long as the initial and final states are guaranteed to satisfy the state of complete energy conversion. Observing Eq. (11), in order to ensure the complete energy conversion of the initial and final states and the existence of $\Delta k$ at the boundary, the following boundary conditions are set:

$$\begin{cases} \theta(0) = 0, & \theta(L) = \pi, \\ \dot{\theta}(0) = -\kappa, & \dot{\theta}(L) = -\kappa, \\ \ddot{\theta}(0) = 0, & \ddot{\theta}(L) = L, \end{cases} \quad (13)$$

In order to satisfy the boundary conditions, the smooth interpolation function is assumed to be of the form of a polynomial ansatz[30], let $\theta(z) = \sum_0^5 a_j z^j$, and the free coefficients $a_j$ can be obtained:

$$\theta(z) = \kappa z - 10(\kappa L - \pi)\left(\frac{z}{L}\right)^3 + 15(\kappa L - \pi)\left(\frac{z}{L}\right)^4 \\ -6(\kappa L - \pi)\left(\frac{z}{L}\right)^5, \quad (14)$$

where $L$ is the length of crystal, and

$$\beta(z) = \arcsin\left(\frac{\dot{\theta}(z)}{\kappa}\right). \quad (15)$$

Thereby, after satisfying the boundary conditions, the coupling coefficient $\kappa$ and phase mismatch $\Delta k$ design still have a great degree of freedom without satisfying the strict adiabatic conditions[5]. Different coupling coefficient corresponds to different interpolation function, that is to say, the evolution of eigenstates of the system is mainly determined by the selection of coupling coefficient This degree of freedom allows us to combine inverse engineering with optimal control to select the most robust state dynamics for demand-based design engineering in the presence of various perturbations and systematic errors while achieving complete energy conversion.

## 3 Optimal schemes against different perturbations

In this section, some perturbations during the evolution of the system are considered, which will influence the robustness of the conversion efficiency. In order to ensure the robustness of the system, we combine LR invariants with optimal theory to propose an optimal scheme for perturbations. System errors shall be considered to consist of perturbations of phase mismatch and coupling coefficient, $\mathbf{H}' = (\eta_\kappa \kappa \sigma_x + \eta_\Delta \sigma_z)/2$ [20], where $\sigma_x$, $\sigma_z$ is the Pauli matrix, $\eta_\kappa, \eta_{\Delta k}$ is the error amplitude, which will be given below for different influence.

Now, considering an ideal, unperturbed original Hamiltonian $H_0$, the system Hamiltonian with perturbation is expressed as $\mathbf{H} = \mathbf{H}_0 + \mathbf{H}'$. Under such conditions, the final state of the conversion efficiency is written[20] as

$$P = |\langle 2|\Psi(L)\rangle|^2 \\ \approx 1 - \frac{1}{4}\left|\int_0^L e^{im}\left(i\eta_\Delta \sin \theta + 2\eta_\kappa \dot{\theta} \sin^2 \theta\right)dz\right|^2, \quad (16)$$

We define coupling coefficient error sensitivity and detuning error sensitivity by perturbation theory, with the consideration of the second-order approximation, respectively

$$q_{\Delta k} = -\frac{1}{2}\frac{\partial^2 P}{\partial \eta_\Delta} = \frac{1}{4}\left|\int_0^L e^{im}\sin\theta\, dz\right|^2, \quad (17)$$

$$q_\kappa = -\frac{1}{2}\frac{\partial^2 P}{\partial \eta_\kappa} = \frac{1}{4}\left|\int_0^L e^{im}2\dot\theta\sin^2\theta\, dz\right|^2, \quad (18)$$

where $m(z) = 2\alpha_+(z) - \beta(z)$.

Then the conversion efficiency can be rewritten as $P \approx 1 - \eta_\kappa^2 q_\kappa - \eta_{\Delta k}^2 q_{\Delta k}$. It can be found that the perturbation is related to the coupling coefficient and phase mismatch, so the perturbation can be nullified by selecting appropriate interpolation function, which is of great significance to maintain the robustness of the system.

### 3.1 $\Delta k$-optimal scheme

First, we discuss the optimal scheme for the perturbation of phase mismatch $\Delta k$. In the process of frequency conversion, the processing error of crystal period is usually regarded as the perturbation of $\Delta k$, i.e., consider perturbation $\mathbf{H}' = \eta_\Delta \sigma_z / 2$, so the error amplitude $\eta_{\Delta k}$ can be rewritten as

$$\eta_{\Delta k} = -\frac{2\pi\delta\Lambda}{\Lambda^2}. \quad (19)$$

and error sensitivity is shown in Eq. (17). To ensure maximum conversion efficiency, the error sensitivity needs to be nullified in the presence of perturbations. Observing Eq. (17), the parameters that determines the value of $q_{\Delta k}$ is the coupling coefficient $\kappa$ and crystal length $L$, and different crystal length $L$ corresponds to different coupling coefficient $\kappa$ of the optimal schemes. By numerical calculation, when $L$=1mm, the coupling coefficient $\kappa$=76.23 cm$^{-1}$ corresponding the minimum error sensitivity $q_{\Delta k}$ =3.6723×10$^{-8}$ m$^2$. This is because there is second-order approximation for error sensitivity, so the result that are approximately equal to 0. By substituting it into Eq. (11), combining the relationship between coupling coefficient and pump amplitude and phase mismatch (the polarization period of PPLN), the pump amplitude $A_2$ as a function of the crystal length $L$ and the polarization period $\Lambda$ as a function of the propagation direction can be expressed as

$$A_2 = \frac{c^2\sqrt{k_1 k_3}}{4\pi\omega_1\omega_3\chi_1}\kappa, \quad (20)$$

$$\Lambda = \frac{2\pi}{\Delta k - k_1 - k_2 + k_3}. \quad (21)$$

In addition, the pump intensity is calculated through $I = 2n_2\sqrt{\varepsilon_0/\mu_0}|A_2|^2$, where $n_2$ is the pump refraction index, $\varepsilon_0$=8.82×10$^{-12}$ F/m and $\mu_0$=4π×10$^{-7}$ H/m. The optimization of pump intensity as shown in red curve of Fig. 1, and the black curve of Fig. 1 represents the pump intensity calculated by the conventional LR invariant. There is an inverse relationship between the pump intensity and the length of the crystal. When $L$=1 mm, quasi-adiabatic optimal scheme for $\Delta k$ needs pump intensity 455 MW/cm$^2$, which is 11 times less than conventional scheme.

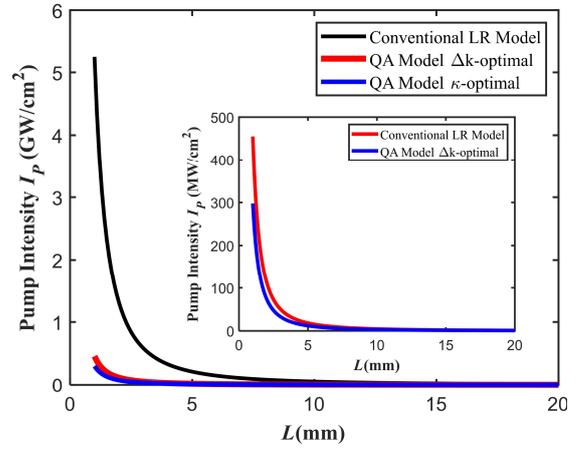

Figure 1 The relationship between pump intensity and crystal length, the black curve is conventional LR model, the red curve is the $\Delta k$-optimal scheme in QA model, and the blue curve is $\kappa$-optimal scheme in QA model. Inset: QA model are magnified

We set the crystal length to 1mm, at which point the optimal polarization period scheme based on inverse engineering for $\Delta k$ perturbation is shown in Fig. 2(a). Then Runge-Kutta method is used to simulate the design with 5 mol % magnesium oxide doped lithium niobate (MgO: LiNbO$_3$) crystal. We choose the pump wavelength is 1.064μm, with the signal wave center wavelength is 3 μm, and the definition of conversion efficiency is

$$\eta = \frac{|A_3(L)|^2}{|A_1(0)|^2}. \quad (22)$$

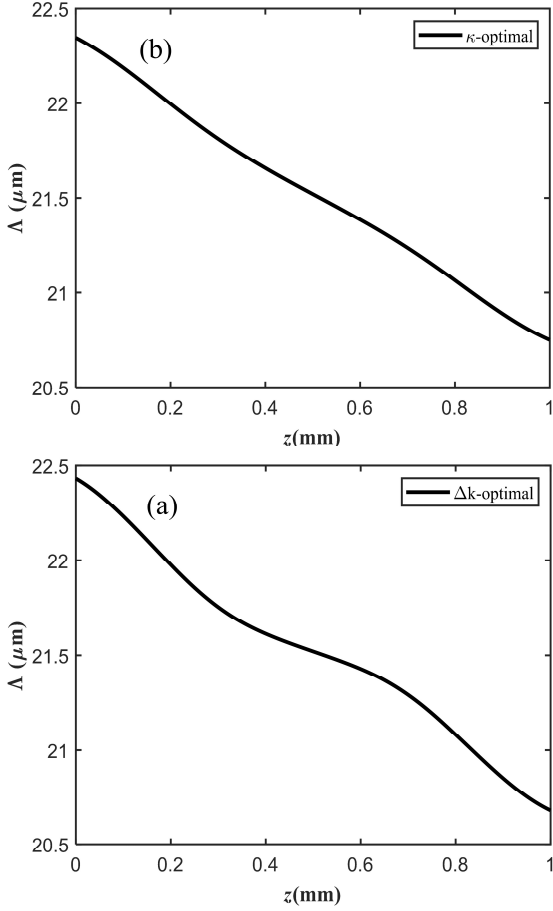

Figure 2 The crystal polarization period is optimal for different perturbations. (a) $\Delta k$-optimal scheme. (b) $\kappa$-optimal scheme.

The interaction between signal light and upconversion light in system evolution is shown in Fig. 3(a). When $L=1$mm, 100% conversion efficiency can be achieved. Then, the relationship between the conversion efficiency and crystal length is researched, as shown in black curve of Fig. 4. which means that for arbitrary length of crystal, complete frequency conversion can be achieved by applying appropriate pump intensity and crystal structure. It is found in the inset that when the crystal length $L<0.1$mm, the conversion efficiency increases gradually from 0 to 100%, and then remains unchanged. Theoretically, the conversion efficiency should always be 100%, which is not realized because there is a minimum limit of crystal length related to the crystal polarization period and the number of periods set. Therefore, it can be approximated that 100% conversion efficiency can be achieved with arbitrary short length under the correct boundary setting. In addition, according to the LZ adiabatic frequency conversion theory, we choose the same coupling coefficient and the extreme value of the phase mismatch $\Delta k$ for simulation comparison with a linear chirp profile of phase mismatch $\Delta k$ under the parameter condition of phase mismatch $\Delta k$ optimization, and find that for the adiabatic theory, only the crystal length $L=2$ mm can achieve highly efficient conversion efficiency with a rapid and violent oscillation asymptotic conversion efficiency profile as shown in the red curve of Fig. 4.

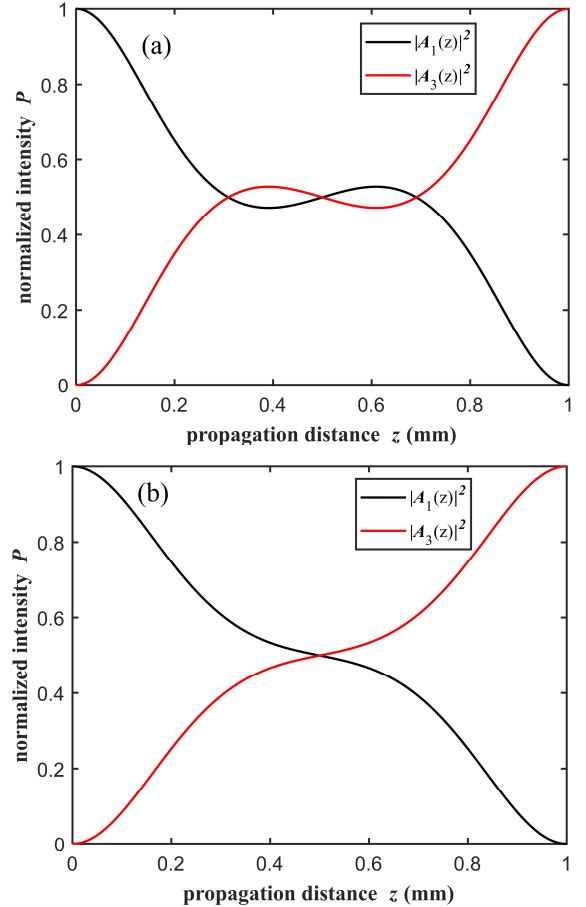

Figure 3 Evolution of the normalized intensity in the crystal along propagation distance z. (a) $\Delta k$-optimal scheme. (b) $\kappa$-optimal scheme.

Then the input wavelength is varied from 2.6 μm to 3.6 μm with a central wavelength of 3 μm. The spectral acceptance bandwidth of signal wave, i.e., full width ratio at half maximum of conversion efficiency, of different crystal lengths is obtained based on this design, as shown in Fig. 5(a). When the crystal length $L=1$ mm, spectral acceptance bandwidth is 393 nm with the guarantee that the conversion efficiency exceeds 90% in the wavelength range of 2.9μm to 3.1μm, but with the increase of the length of the crystal, the bandwidth reduced to 40nm in $L$=20 mm. This is remarkable since these bandwidths are generally inversely proportional to the length of bulk nonlinear crystal. The black dotted lines and red dotted line represent the conversion bandwidth of the conventional invariant method using the detuning optimal scheme and LZ adiabatic method respectively when the crystal length $L$=1 mm. By contrast, it is found that the spectral bandwidth of our quasi-adiabatic method is 28 nm smaller than that of the conventional LR invariant method under the condition of the same crystal size. At this time, although the bandwidth of LZ adiabatic method is 339 nm, its maximum conversion efficiency is only 64%, this's because our scheme doesn't need to satisfy strict adiabatic conditions.

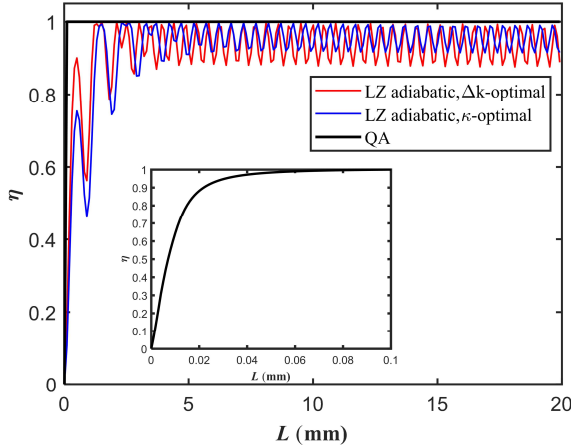

Figure 4 The conversion efficiency of the two adiabatic methods varies with crystal length. Red curve is the LZ adiabatic scheme with the parameters of $\Delta k$-optimal scheme, blue curve is the LZ adiabatic scheme with the parameters of $\kappa$-optimal scheme and the black curve is QA model. Inset: conversion efficiency when crystal length less than 0.1mm.

It is found that the conversion efficiency profile is not smooth with some spectral ripple. This is due to the interaction with large phase mismatch at the beginning and end, which means that when light waves enter or leave the nonlinear medium, there are abrupt changes in system parameters. It's essentially wave coupling on or off, phase mismatch is finite when light waves enter or leave nonlinear media, so that the system doesn't evolve adiabatically between the eigenstates. Instead, it excites the superposition of these eigenstates, and spectral ripples are the manifestations of mutual interference between these eigenstates. This is consistent with the bandwidth profile produced by an adiabatic crystal without an apodization procedure[5, 31].

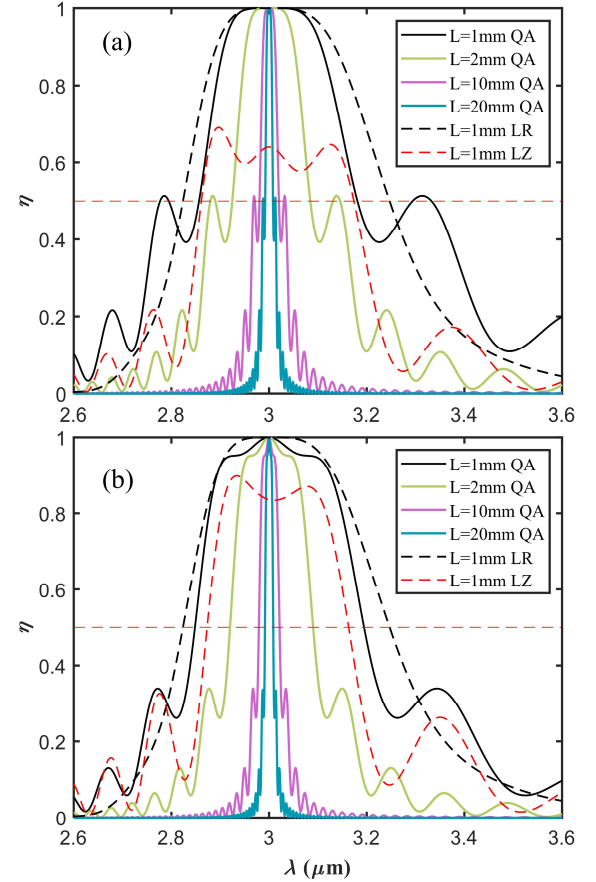

Figure 5 The conversion efficiency and spectral acceptance bandwidth of different crystal lengths with different optimization for perturbations. (a) $\Delta k$-optimal scheme. (b) $\kappa$-optimal scheme.

Next, we consider the effect of perturbation on conversion efficiency, which is shown in the Fig. 6, where the black curves in a, b are the change of conversion efficiency in the geometrical imperfection (e.g., crystal polarization period processing error) and pump intensity variation respectively and the black dotted line shows the optimal results of the traditional LR invariant method, and the blue dotted line shows the robustness of the LZ adiabatic method. For the polarization period, when the processing error is between -20% to +20%, the frequency conversion can be realized, and it has strong robustness with the almost 100% conversion efficiency between -1% to 1% in the inset. The profile of this curve isn't symmetrical, because when the crystal period increases or decreases by the same value, the change of phase mismatch $\Delta k$ is not the same, and the system has higher tolerance when the crystal period increases. And LZ adiabatic method has a lower robustness against this perturbation. It's because that LZ adiabatic method has a strict limit on the rate of change of phase mismatch, but QA model doesn't have. When the pump intensity variation is in the range -25% to +10%, it has a high tolerance over 80%, as well as when the actual pump intensity is less than the set, there is a higher tolerance at the condition of small variation. Overall, the result shows the desired robustness against two kinds of perturbations at a short crystal length, as a result of our optimal scheme.

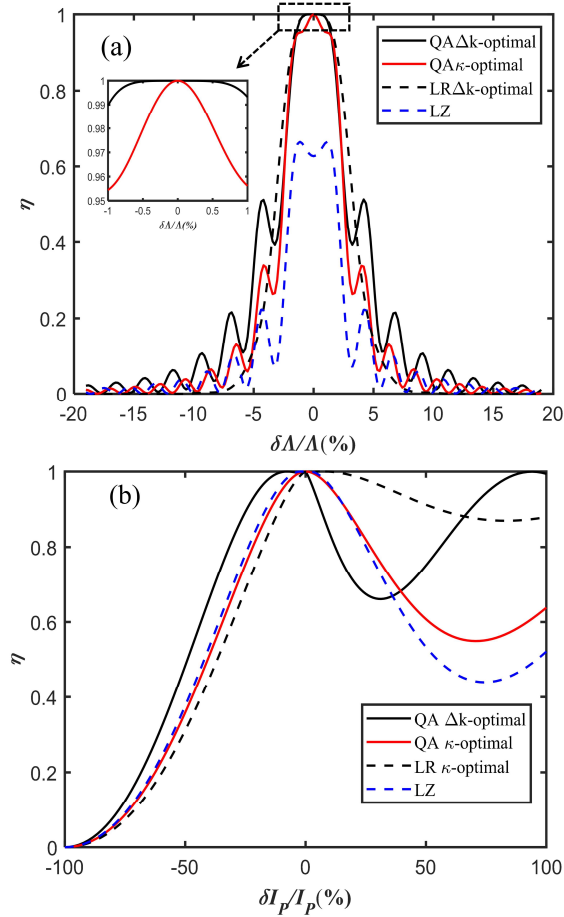

Figure 6 Tolerance of different perturbations in different optimal schemes. Black curve is a univariate model $\kappa$-optimal scheme and red curve is $\Delta k$-optimal scheme. (a) Crystal polarization period processing error. (b) Pump intensity variation.

### 3.2 $\kappa$-optimal scheme

Next, we apply the optimal protocol for robustness against the perturbation of coupling coefficient $\kappa$ caused by pump intensity variation, i.e., $\mathbf{H}_\kappa = \kappa \eta_\kappa \sigma_x / 2$, with the error amplitude $\eta_\kappa$ can be rewritten as

$$\eta_\kappa = \frac{4\pi \omega_1 \omega_3}{c^2 \sqrt{k_1 k_3}} \chi_1 \delta A_2, \qquad (23)$$

which is obtained from Eqs. (6), (14), and (18) to design a crystal at a length of $L$ =1 mm, and the error sensitivity is shown in Eq. (18). The coupling coefficient corresponding to the perturbation optimal solution, i.e., $\kappa$=61.33 cm$^{-1}$, $q_\kappa$=1.2328×10$^{-6}$ m$^2$, which can be substituted into Eqs. (20) and (21) to obtain the optimal polarization period scheme for the perturbation of coupling coefficient, as shown in Fig. 2(b). This optimal scheme is 17 times lower than the pump intensity required by conventional theory, as shown in the blue curve in Fig. 1. Again, we take the same parameters as the previous section and calculate the other parameters for the frequency conversion. The interaction between signal light and upconversion light in system evolution is shown in Fig. 3(b). Under this design, complete energy transfer can also be achieved when $L$=1 mm. The evolution process is different and our design is sensitive to changes in crystal length when the other parameters is determined. The blue curve in Fig. 4 represents the LZ adiabatic evolution under the parameter condition of coupling optimization, which requires $L \geq $ 2mm to achieve high conversion efficiency.

The bandwidth of our QA model in Fig. 5(b) is 345nm when $L$=1 mm, which is larger than LZ adiabatic method, and when $L$=20 mm, the bandwidth is 17 nm. Moreover, the conversion efficiency profile has a fast attenuation near the central wavelength, because wavelength changes mainly affect phase mismatch, so the optimization of phase mismatch $\Delta k$ has greater tolerance against polarization period processing error.

The effect of actual parameter variation is considered, the red curves in Fig. 6 represents the relationship between pump intensity variation, polarization period processing error and conversion efficiency based on the $\kappa$-optimal scheme respectively. Since the polarization period only affects the phase mismatch, the optimal scheme for the perturbation of coupling coefficient $\kappa$ is relatively poor in the robustness of the polarization period processing error. Nevertheless, a 95% conversion efficiency can still be achieved with an error of -1% to 1%.

When the pump intensity varies between -25% and 25%, conversion efficiency is greater than 80%, and if the actual pump intensity ratio is larger than the set, there is a higher tolerance at the condition of large variation. For the optimization of pump light intensity variation, the conversion efficiency of $\kappa$ optimal scheme is more robust than that of $\Delta k$, because the pump intensity variation mainly affects the coupling coefficient $\kappa$. As well as the conventional invariant method has an advantage because it optimizes both $\theta$ and $\beta$ variables simultaneously. When the variation is small, the quasi-adiabatic method is nearly as robust as the conventional method, and all of these methods is more robust tan the LZ adiabatic method.

The following shows the relationship between signal light intensity and conversion efficiency. When the approximation of undepleted pump is not satisfied, as shown in the Fig. 7, for the two optimal schemes, when the ratio of the signal light amplitude to the pump light amplitude is less than 0.1 and 0.05, the conversion efficiency remains almost unchanged at 100%, and then the conversion efficiency gradually decreases. When the intensity of signal wave and pump wave is equal, the conversion efficiency is 77% and 70%, respectively. Additionally, when the signal intensity is less than $\mu W/cm^2$, as shown in the inset of Fig. 7, the conversion efficiency contains almost 100%. This is due to when the signal light intensity is very weak, the undepleted pumping approximation is perfectly satisfied. According to our theory, the final state of the evolution of the system is set as a complete conversion of up-conversion light, as shown in Eq. (12). Therefore, as long as the boundary conditions of Eq. (13) are satisfied, the coupling coefficient $\kappa$ and phase mismatch $\Delta k$ obtained by inverse engineering can ensure that the conversion efficiency is always close to 100%. This shows that our method is more suitable for the frequency conversion process of small signals.

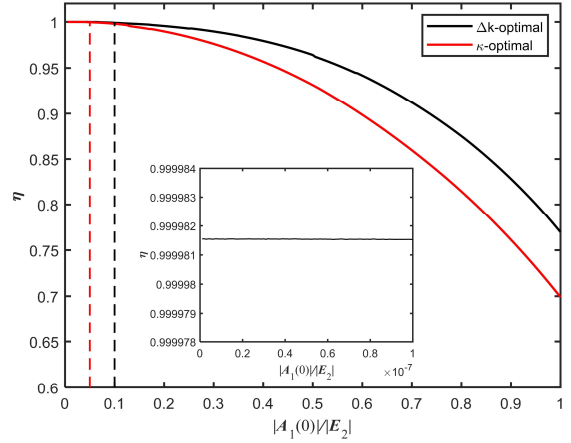

Figure 7 The relation between different signal light amplitudes and conversion efficiency. Inset: Conversion efficiency of signal light with low intensity

## 4  Discussion and conclusion

In this paper, based on the analogy between the wave equations of light propagation in nonlinear crystals and the dynamics of two-level atomic systems, we propose a quasi-adiabatic single control parameter model based on the STA Lewis-Riesenfeld invariant and optimal theory. The model shows that when the initial and final states of the system are given, by changing the values of single control parameters, the interpolation functions of different profiles are selected, and the robustness of the system changes correspondingly. By using optimal theory, the optimal solution of single control parameter is obtained, and the most robust evolution path of the system under different perturbations is obtained. Under the setting of the single control parameter, i.e., coupling coefficient $\kappa$ is constant, and considering the effects of two kinds of perturbations, i.e., coupling coefficient and phase mismatch, the nonlinear crystal structure that is robust to different perturbations is obtained by inverse engineering of phase mismatch $\Delta k$. This method can accelerate adiabatic techniques for frequency conversion to achieve almost 100% efficient nonlinear frequency conversion at any crystal length, which means complete energy conversion can be achieved in shorter distances. Furthermore, by designing the $\kappa$-optimal and $\Delta k$-optimal scheme, when the small crystal size $L$=1 mm, the spectral acceptance bandwidth of 393 nm and 345 nm is achieved in the mid-infrared signal light from 2.6 μm to 3.6 μm respectively. In addition, the pump intensity of the $\kappa$-optimal and $\Delta$k-optimal scheme is 17 and 11 times lower than that of the conventional scheme respectively, but this is at the cost of sacrificing little robustness of the system. Afterward the robustness of the $\Delta$k optimization against crystal period variation as well as $\kappa$ optimization against pump intensity variation is discussed. Obviously, the emphasis of the two optimal schemes is different, we need to take a tradeoff in the target parameters to ensure that one of the indicator parameters is better, which depends on the importance of the parameter in the system. Even so, our method is superior to LZ adiabatic theory under the condition of the same parameters, and if the error sensitive coefficient is discussed to a higher order term, more robust results can be obtained. Finally, the relationship between signal light intensity and conversion efficiency shows that when the signal light intensity is less than 1% of the pump intensity, high conversion efficiency process can be theoretically realized, because our theory is based on the approximation of undepleted pump, which can convert the weak signal light well.

In conclusion, we confirm numerically that the invariant-based inverse engineering approach manages to efficiently convert the optical beam frequency with shorter propagation distance, as compared to LZ adiabatic scheme. Combining with perturbation theory, different nonlinear crystal structures can be designed with more freedom by using quasi-adiabatic model that are robust against perturbations due to different variables. This method can be applied to design efficient optical converters to upconvert weak mid-infrared signals, we believe that this technique provides a new design idea in highly efficient frequency conversion process.

## 5 Acknowledgments

This work is supported by the National Natural Science Foundation of China (NFSC) (61975232 and 61775234).